\documentclass[12pt]{article}
\def\s{\sigma}

\def\bpsi{\bar{\psi}}
\def\btheta{\bar{\theta}}
\def\bchi{\bar{\chi}}
\def\baeta{\bar{\eta}}
\def\bQ{\bar{Q}}

\def\bepsilon{\bar{\epsilon}}

\def\Q1{Q^{1}}
\def\Q2{Q^{2}}

\def\CD{\mathcal{D}}
\def\CW{\mathcal{W}}
\def\dfrac#1#2{{\displaystyle\frac{#1}{#2}}}
\def\cfrac#1#2{\dfrac{\mathstrut #1}{#2}}
\newcommand{\norsl}{\normalsize\sl}
\newcommand{\norsc}{\normalsize\sc}
\begin{document}
\begin{titlepage}

\title{
Nonlinear Realization of Partially Broken $N=2$ \\
AdS Supersymmetry in Two and Three Dimensions   
}

\author{ \\
\norsc  Masayuki SANO\thanks{e-mail:masayuki@phys.h.kyoto-u.ac.jp}  
\thanks{Now at Graduate School of Informatics, Kyoto University}  \\
\norsl  Graduate School of Human and \\
\norsl  Environmental Studies, Kyoto University\\
\norsl  Kyoto 606-8501, JAPAN\\
\\
and
\\
\\
\norsc       Tsuneo UEMATSU\thanks{e-mail:uematsu@phys.h.kyoto-u.ac.jp}
\\
\norsl  Department of Fundamental Sciences\\
\norsl  FIHS, Kyoto University\\
\norsl  Kyoto 606-8501, JAPAN\\
\\
}

\date{}

\maketitle

\begin{abstract}
{\normalsize
We investigate the nonlinear realization of partially broken $N$=2 
global supersymmetry in the $D=2$ and $3$ anti de-Sitter (AdS) space. 
We particularly study Nambu-Goldstone
degrees of freedom for the 
$N$=2 AdS supersymmetry partially broken
down to $N$=1 AdS supersymmetry, where we observe a NG fermion of the broken
supersymmetry and a NG boson of the internal symmetry which form a NG 
multiplet.  Based on the nonlinear realization method, 
we construct a superspace formalism for 2 and 3 dimensional AdS space and 
evaluate the covariant derivatives and supervielbeins
for the AdS superspace. Finally we obtain the nonlinear transformation laws and
the lowest order of effective Lagrangians.
}
\end{abstract}

\begin{picture}(5,2)(-310,-660)
\put(35,-105){KUCP-151}
\put(35,-120){December 2000}
\end{picture}

\vspace{2cm}

\thispagestyle{empty}
\end{titlepage}
\setcounter{page}{1}
\baselineskip 20pt
%
\vspace{0.3cm}
\leftline{\large\bf 1. Introduction}
\vspace{0.3cm}

The $N$=1 supersymmetry in the minimal supersymmetric standard model 
is thought to be realized as the low energy limit of a more
fundamental theory formulated in higher dimensions with 4-dimensional
$N$-extended supersymmetry, 
which has to be spontaneously broken to $N=1$. 
In such a context, partial breaking of extended supersymmetry has been 
studied in the literatures \cite{HLP,APT,FGP,BG1,BG2} 
in the last several years.

Hughes, Liu and Polchinski \cite{HLP} first pointed out the possibility of the
partial supersymmetry breaking
and constructed an example arising from 
the 4-dimensional supermembrane solution 
of 6-dimensional supersymmetric gauge theory.
While, Antoniadis, 
Partouche and Taylor \cite{APT} introduced the electric and magnetic 
Fayet-Iliopoulos terms  in the $N$=2 gauge theory of abelian vector multiplet 
and have shown 
the spontaneous breaking of $N$=2 to $N$=1.
This partial breaking induced by the Fayet-Iliopoulos terms 
has also been obtained by taking the flat limit of the $N$=2 supergravity 
theories \cite{FGP}.

On the other hand, 
Bagger and Galperin \cite{BG1,BG2} have studied the 
nonlinear realization of $N$=2 supersymmetry partially broken down to $N$=1 
supersymmetry in $D$=4 flat space \cite{BW,SW}.
They obtained the Nambu-Goldstone (NG) multiplet both for the cases; chiral
multiplet \cite{BG1} and vector-multiplet \cite{BG2}, and discussed 
the nonlinear transformation laws as well as the low-energy effective 
Lagrangians \cite{B}.
 
Now the partial breaking of supersymmetry in flat space can be extended to
the anti-de Sitter (AdS) space, which has recently attracted much attention 
in the context of AdS/CFT correspondence as well as brane-world scenario.
By unHiggsing the $N$=1 massive spin-3/2
multiplet, Altendorfer and Bagger studied the partial breaking of the 
$N=2$ AdS supersymmetry $OSp(2,4)$ \cite{AlB}.
For AdS background, Zumino and Deser \cite{Z,DZ} 
have investigated the nonlinear realization of $N=1$ AdS supersymmetry.
They have found that there appears the ``mass'' term of the NG fermion 
which is proportional to the radius of AdS space in the effective Lagrangian 
and considered its relation to the super-Higgs effect.

Here in this paper, we shall investigate the nonlinear realization of
$D$=2 and 3, $N$=2 global AdS supersymmetry. We study the
spontaneous breaking of this symmetry down to $N$=1 AdS
supersymmetry using the AdS superfield method.
We find that there appears only one $N$=1 NG multiplet which
consists of a NG fermion of the broken second 
supersymmetry and a NG boson of the broken internal symmetry.
We next consider the nonlinear transformation laws and then
construct the effective Lagrangian up to the second order of
fields.
It is intriguing to examine whether a peculiar value of ``mass'' 
term would emerge as $N$=1 case studied 
in \cite{Z}.

\vspace{0.3cm}
\leftline{\large\bf 2. Algebras and
 Partial Breaking of AdS supersymmetry}
\vspace{0.3cm}

In this section, we consider the $N$=2 AdS superalgebra and its
spontaneous breaking. We then determine the relations among the NG
fields associated with the broken generators.

\vspace{0.1cm}
\leftline{\bf 2.1.  $D$=2 and 3 AdS superalgebras}
\vspace{0.1cm}

The two and three dimensional AdS superalgebras \cite{Keck} are the algebras 
obtained by adding $N$ spinorial generators to the bosonic AdS algebras,
and are given as following (see Appendix for the notations):
\[
 [P_{\mu}, P_{\nu}] = -im^{2}M_{\mu\nu},\ \ 
 \protect{[}P_{\mu}, M_{\rho\sigma}] = i\eta_{\mu\rho}P_{\sigma}
  - i\eta_{\mu\sigma}P_{\rho},
\]
\[
 \protect[M_{\mu\nu}, M_{\rho\sigma}] = -i\eta_{\mu\rho}M_{\nu\sigma}
  - i\eta_{\nu\sigma}M_{\mu\rho} + i\eta_{\mu\sigma}M_{\nu\rho}
  + i\eta_{\nu\rho}M_{\mu\sigma},
\]
\[
  [Q^{i \alpha}, P_{\mu}]
  = \frac{m}{2}(\gamma_{\mu}Q^{i})_{\alpha},\ \ 
 [Q^{i}_{\alpha}, M_{\mu\nu}]
  = \frac{1}{2}(\sigma_{\mu\nu}Q^{i})_{\alpha}, \quad (i=1,2,\cdots, N),
\]
\begin{eqnarray}
 \{Q^{i}_{\alpha}, \bQ^{j \beta}\}
  &=& m\delta^{ij}(\s^{\mu\nu})_{\alpha}^{\>\>\beta}M_{\mu\nu}
  + 2\delta^{ij}(\gamma^{\mu})_{\alpha}^{\>\>\beta}P_{\mu}
  + i \delta^{\beta}_{\alpha}T^{ij} (D=2), \label{eq:alg}\\
  &=& m\delta^{ij}(\s^{\mu\nu})_{\alpha}^{\>\>\beta}M_{\mu\nu}
  + 2\delta^{ij}(\gamma^{\mu})_{\alpha}^{\>\>\beta}P_{\mu}
  + 2i \delta^{\beta}_{\alpha}T^{ij} (D=3), \nonumber
\end{eqnarray}
\[
 [T^{ij}, Q_{\alpha}^{k}]
  = im\delta^{ik}Q_{\alpha}^{j} - im\delta^{jk}Q_{\alpha}^{i},
\]
\[
 [T^{ij}, T^{kl}] = -i\delta^{jk}T^{il} + i\delta^{ik}T^{jl}
 + i\delta^{jl}T^{ik} - i\delta^{il}T^{jk},
\]
where $T^{ij}$'s are $SO(N)$ generators of the internal symmetry of
the supercharges and $\sigma^{\mu\nu}=\frac{i}{2}[\gamma^{\mu},\gamma^{\nu}]
= i\epsilon^{\mu\nu}\gamma_{5}\,(D=2), -\epsilon^{\mu\nu\rho}\gamma_{\rho}\,(D=3)$. 
We can recover $SO(d-1,2)$ algebra which is the isometry
of the $d$ dimensional AdS space, given by hyperboloid 
$\eta_{AB}y^{A}y^{B} = 1/m^{2}$ with $\eta_{AB} = (\eta_{\mu\nu},+1)$ ,
$\eta^{\mu\nu}=(+1,-1,\ldots -1)$,
by rewriting 
the generators of pseudo-translation $P_{\mu}$ as
 $P_{\mu} \equiv m M_{\mu d}$, where $m$ is the reciprocal 
of the radius of AdS space.
Note that by the Majorana properties of the supercharges $Q_\alpha^i$,
the internal symmetry becomes $SO(N)$. 
For $N$=2, there is only one internal symmetry generators 
$T^{12}$ which we shall denote by $T$ in the following.
This algebra is contracted to
the super-Poincar{\' e} algebra in the flat limit $m \rightarrow 
0$.

\newpage

\vspace{0.1cm}
\leftline{\bf 2.2.  Nambu-Goldstone multiplet}
\vspace{0.1cm}

Now we identify the $N$=1 NG multiplets for the 
partially broken $N$=2 AdS superalgebra.
In our case where $N$=2 AdS superalgebra is spontaneously broken 
down to $N$=1 AdS superalgebra, we have the broken generators
 $Q^{2}_{\alpha}$ and $T$. 
We assign a fermion $\psi_{\alpha}$ and a boson $\phi$ 
to the corresponding NG particles, respectively.
Because of the unbroken $N=1$ supersymmetry, these particles
must have its own superpartners. 
But in this case 
it turns out that $\psi$ and $\phi$ 
are the superpartners of each other, which can be seen from
the following argument \cite{KU,GU}.

Let us consider the Jacobi identity for $Q^{1}$, $T$ and $\psi$ given
by 
\begin{equation}
 \{[Q^{1}_{\alpha},T], \psi\} - \{[T, \psi], Q^{1}_{\alpha}\}
  -[\{\psi, Q^{1}_{\alpha}\}, T] = 0.
	\label{jacobipsi}
\end{equation}
Taking the vacuum expectation value of (\ref{jacobipsi}) we have
\begin{equation}
 \langle 0|[ \{\psi, Q^{1}_{\alpha}\}, T]|0\rangle =
  -im\langle 0| \{Q^{2}_{\alpha}, \psi\}|0\rangle \neq 0,
\end{equation}
where we used the fact $Q^{1}|0\rangle = 0$ and
 $\langle 0| \{Q^{2}_{\alpha}, \psi\}|0\rangle \neq 0$.
Since $\langle 0|[\phi, T]|0\rangle \neq 0$ for $\phi$, the NG boson 
of $T$, this means that
\begin{equation}
 \phi \sim \{\psi, Q^{1}_{\alpha}\},
\end{equation}
i.e. the supersymmetry transformation of the NG fermion of the second
supersymmetry, $\psi$, is related to the NG boson of the
internal symmetry, $\phi$.

\vspace{0.3cm}
\leftline{\large\bf 3. Coset Construction and AdS Superspace}
\vspace{0.3cm}

When we apply the framework of the nonlinear realization \cite{CWZOg} to the
analysis of the partially broken AdS supersymmetry, we must
represent the NG field as a superfield. In this section we
formulate the AdS superfield using a group theoretical method \cite{IS}.

For $D=2$ and $3$, the $N=1$ scalar multiplet
can be represented as a real scalar function $\Phi(x^{\mu},\theta_{\alpha})$ 
\begin{equation}
 \Phi(x,\theta) = A(x) + \btheta\psi(x) + \frac{1}{2}\btheta\theta F(x),
  \label{2dflatphi}
\end{equation}
where $A(x)$, $\psi(x)$ and $F(x)$ 
are a real scalar, a Majorana
spinor and an auxiliary scalar, respectively.
Here we note that the $\Phi$ represents an irreducible $N$=1 scalar multiplet.

First, we consider supercovariant derivatives for AdS superfield.
The AdS superspace is given as the coset space of 
the AdS algebra (\ref{eq:alg}) represented by the coset representive
\begin{equation}
 L(x,\theta) = e^{iz^{\mu}(x)P_{\mu}}e^{i\btheta^{\alpha} Q_{\alpha}},
\end{equation}
where $z^{\mu}(x) = \frac{2}{mx}\tan ^{-1}\frac{mx}{2}{x^{\mu}}, 
x=(x^{\mu}x_{\mu})^{1/2}$.
Here $x^{\mu}$'s are the coordinates of AdS space and
$\theta^{\alpha}$'s are the fermionic coordinates. In these coordinates, the
zweibein (dreibein) of the AdS space and the spin connection take the form
\begin{equation}
 e^{a}_{\mu} =
 \cfrac{\delta^{a}_{\mu}}{1+\frac{m^{2}x^{2}}{4}},\ \ 
 \omega^{ab}_{\mu} = \frac{m^{2}}{2}
   \cfrac{x^{a}\delta^{b}_{\mu}-x^{b}\delta^{a}_{\mu}}
	{1+\frac{m^{2}x^{2}}{4}}.
\end{equation}

We calculate the Cartan differential 1-form as \cite{Z}
\begin{equation}
 L^{-1}dL \equiv iDx^{a}P_{a}
  + iD\btheta^{\alpha}Q_{\alpha}
  + \frac{i}{2}\mathit{\Delta\Omega}^{ab}M_{ab},
\end{equation}
where
\begin{eqnarray}
 Dx^{a} &=& \left(1+\frac{m}{2}\btheta\theta\right)\nabla x^{a}
      + i(d\btheta\gamma^{a}\theta) \quad (D=2)\nonumber \\
	&=& \left(1+\frac{m}{2}\btheta\theta\right)\nabla x^{a}
      - \frac{1}{4}\epsilon^{bca}\mathit{\Delta}\omega^{bc}(\btheta\theta)
      + i(d\btheta\gamma^{a}\theta) \quad (D=3),\nonumber \\
 D\btheta^{\alpha} &=& 
      -i\frac{m}{2}\nabla x^{a}(\btheta\gamma_{a})^{\alpha}
      + \frac{1}{4}\mathit{\Delta}\omega^{ab}\epsilon_{ab}
      (\btheta\gamma_{5})^{\alpha} +
      \left(1-\frac{m}{4}(\btheta\theta)\right)d\btheta^{\alpha} \quad (D=2)
	  \label{ospcartanresult} \\
      &=& 
	 -i\frac{m}{2}\nabla x^{a}(\btheta\gamma_{a})^{\alpha}
          + \frac{i}{4}\mathit{\Delta}\omega^{ab}\epsilon_{abc}
                (\btheta\gamma^{c})^{\alpha} + 
	\left(1-\frac{m}{2}(\btheta\theta)\right)d\btheta^{\alpha} \quad (D=3),
	\nonumber \\
 \frac{1}{2}\mathit{\Delta \Omega}^{ab}
	 &=& 
   \left(1+\frac{m}{2}\btheta\theta\right)\frac{1}{2}\mathit{\Delta}\omega^{ab}
     - \frac{m}{2}\epsilon^{ab}(d\btheta\gamma_{5}\theta) \quad (D=2)
	 \nonumber \\
	&=&
   \left(1+\frac{m}{2}\btheta\theta\right)\frac{1}{2}\mathit{\Delta}\omega^{ab}
     - \frac{1}{4}m^{2}\nabla x_{c}\epsilon^{cab}(\btheta\theta)
     - i\frac{m}{2}\epsilon^{abc}(d\btheta\gamma_{c}\theta) \quad (D=3), 
	\nonumber
\end{eqnarray}
where we use the notations $\nabla x^{a} =
e^{a}_{\mu}dx^{\mu}$,
$\mathit{\Delta}\omega^{ab} = - \omega^{ab}_{\mu}dx^{\mu}$.

The superzweibein (dreibein) $W_{M}^{\>\>A}$ is defined as
\begin{equation}
 Dz^{A} = dz^{M}W_{M}^{\>\>A},\ \  Dz^{A}=(Dx^{a}, D\btheta^{\alpha})
  ,\ \ dz^{M}=(dx^{\mu}, d\theta^{\beta}).
\end{equation}
\relax From the formula of Cartan form (\ref{ospcartanresult}), we
obtain $W_{M}^{\>\>A}$ and its inverse becomes
\begin{eqnarray}
 (W^{-1})_{A}^{\>\>M} 
 &=& \left(\begin{array}{@{\,}cc@{\,}}
      (W^{-1})_{a}^{\>\>\mu} & (W^{-1})_{a}^{\>\>\beta} \\
      (W^{-1})_{\alpha}^{\>\>\mu} & (W^{-1})_{\alpha}^{\>\>\beta}
      \end{array}\right) \nonumber \\
 &=& \left(
  \begin{array}{@{\,}cc@{\,}}
   e^{\mu}_{a} & -\cfrac{im}{2}(\btheta\gamma_{a})^{\beta}
                 + \cfrac{1}{2}\omega_{a}(\btheta\gamma_{5})^{\beta} \\
   -ie^{\mu}_{a}(\gamma^{a}\theta)_{\alpha}
               & \left(1-\cfrac{m}{4}(\btheta\theta)\right)
	       \delta^{\beta}_{\alpha}
	       - \cfrac{i}{4}(\btheta\theta)\omega_{a}\epsilon^{ad}
	                                (\gamma_{d})^{\beta}_{\alpha}
  \end{array}\right)\quad (D=2) \\
 &=& \left(
  \begin{array}{@{\,}cc@{\,}}
   e^{\mu}_{a} & \cfrac{im}{2}(\btheta\gamma_{a})^{\beta}
                 + \cfrac{i}{2}\omega_{ac}(\btheta\gamma^{c})^{\beta} \\
   -ie^{\mu}_{a}(\gamma^{a}\theta)_{\alpha}
               & \left(1-\cfrac{m}{4}(\btheta\theta)\right)
	       \delta^{\beta}_{\alpha}
     			- \cfrac{1}{4}(\btheta\theta)\omega_{\mu}^{c}
	                       (\gamma_{c}\gamma^{\mu})^{\beta}_{\alpha}
  \end{array}\right)\quad (D=3), \nonumber
  \label{ospzweibeinresult}
\end{eqnarray}
where we define
 $\omega_{\mu}=\frac{1}{2}\omega_{\mu}^{ab}\epsilon_{ab}\,(D=2)$,
 $\omega_{\mu a}=\frac{1}{2}\omega_{\mu}^{bc}\epsilon_{abc}\,(D=3)$.
Then we introduce the supercovariant derivative of general superfield $\Phi$
 which transform linearly under the AdS supersymmetry as follows:
\begin{equation}
 D_{A}\Phi = \frac{D\Phi}{Dz^{A}} = (W^{-1})_{A}^{\>\>M}\frac{D\Phi}{dz^{M}},
 \quad
 D\Phi \equiv d\Phi + \frac{i}{2}\mathit{\Delta}\omega^{bc}\Sigma_{bc} 
 \Phi.
\end{equation}
where $\Sigma_{bc}$ is the matrix representation of the Lorentz group. Hence,
\begin{eqnarray}
 \cfrac{D\Phi}{Dx^{a}}
 &=& \nabla_{a}\Phi 
  + \cfrac{im}{2}\left(\btheta\gamma_{a}
		  \cfrac{\partial\Phi}{\partial\btheta}\right)
   +\cfrac{1}{2}\omega_{a}\left(\btheta\gamma_{5}
			   \cfrac{\partial\Phi}{\partial\btheta}\right)
	\quad (D=2) \nonumber \\
    \label{ospcvx}
 &=& \nabla_{a}\Phi 
  + \cfrac{im}{2}\left(\btheta\gamma_{a}
		  \cfrac{\partial\Phi}{\partial\btheta}\right)
   +\cfrac{i}{2}\omega_{ac}\left(\btheta\gamma^{c}
			   \cfrac{\partial\Phi}{\partial\btheta}\right)
   -\frac{im}{4}(\btheta\theta)\omega_{a}^{bc}\Sigma_{bc}\Phi 
	\\
 & & {} \hspace{-2cm}
   -\frac{im^{2}}{4}\epsilon^{abc}(\btheta\theta)\Sigma_{bc}\Phi
   +\frac{im^{2}}{4}\epsilon^{dbc}(\btheta\gamma_{a}\gamma_{c}\theta)
	\Sigma_{db}\Phi
   + \frac{im}{4}\omega_{ae}(\btheta\gamma^{e}\gamma_{c}\theta)
	\epsilon^{dbc}\Sigma_{db}\Phi \quad (D=3), \nonumber \\
 \cfrac{D\Phi}{D\btheta^{\alpha}}
  &=& -i(\gamma^{a}\theta)_{\alpha}\cfrac{D\Phi}{D x^{a}}
  + \left(1 + \cfrac{m}{4}(\btheta\theta)\right)
  \cfrac{\partial\Phi}{\partial\btheta^{\alpha}}
  -\cfrac{im}{2}\epsilon^{bc}(\gamma_{5}\theta)_{\alpha}
  \mathit{\Sigma_{bc}}\Phi \quad (D=2)\nonumber \\
  &=& -i(\gamma^{a}\theta)_{\alpha}\cfrac{D\Phi}{D x^{a}}
  + \left(1 + \cfrac{m}{2}(\btheta\theta)\right)
  \cfrac{\partial\Phi}{\partial\btheta^{\alpha}}
  +\cfrac{m}{2}\epsilon^{bca}(\gamma_{a}\theta)_{\alpha}\Sigma_{bc}\Phi
  \quad (D=3), \label{ospcvspinor}
\end{eqnarray}
where $\nabla_{a}$ is the AdS covariant derivative. 
As $m \rightarrow 0$, these operators 
are reduced to the usual supercovariant derivatives in the flat superspace.

Next we consider the transformation laws. Under the
multiplication of a group element $g$ of AdS supersymmetry
group, the left-coset $L(x,\theta)$ transforms as \cite{Z}
\begin{equation}
 L \rightarrow gL=L'h \quad,
 \label{genmulti}
\end{equation}
$h$ is a group element of Lorentz group. 
In particular, for an infinitesimal transformation, we take $g = 1 +
i\bepsilon^{\alpha}Q^{1}_{\alpha}$, $L^{\prime} = L + \delta L$,
$h = 1 + i\delta u^{\mu\nu}M_{\mu\nu}$ and (\ref{genmulti}) becomes
\begin{equation}
 e^{-i\btheta Q}e^{-iz(x)\cdot P}(i\bepsilon Q)e^{iz(x)\cdot P}e^{i\btheta Q}
  - e^{-i\btheta Q}e^{-iz(x)\cdot P}\delta(e^{iz(x)\cdot P}e^{i\btheta Q}) 
  = i\delta u^{\mu\nu}M_{\mu\nu}.
  \label{osptrformula}
\end{equation}
To get the transformation laws of $x^{\mu}$ and $\theta^{\alpha}$
 we must evaluate the left hand side of (\ref{osptrformula})
and set the coefficients of $P_{a}$ and $Q_{\alpha}$ to zero.
Then we obtain the equations of $\delta x^{\mu}$ and 
$\delta \btheta^{\alpha}$, which leads to the transformation laws
\begin{eqnarray}
 \delta x^{\mu} &=& i(\baeta\gamma^{\mu}\theta)  \quad (D=2, 3), \nonumber \\
 \delta \btheta^{\alpha} &=& \baeta^{\alpha}
  - \frac{i}{4}(\btheta\theta)\omega_{a}
    (\baeta\gamma^{a}\gamma_{5})^{\alpha} \quad (D=2) \\
 &=& \left(1-\frac{m}{4}(\btheta\theta)\right) \baeta^{\alpha}
    - \frac{i}{4}(\btheta\theta)\omega_{ab}
             (\baeta\sigma^{ab})^{\alpha} \quad (D=3). \nonumber
\end{eqnarray}
Here we difine $\eta_{\alpha}$ by
\begin{equation}
e^{-iz(x)\cdot P}(i\bepsilon^{\alpha} Q_{\alpha})e^{iz(x)\cdot P}
	 \equiv i\baeta^{\alpha} Q_{\alpha}.
\end{equation}

For the transformation laws, the parameter $\epsilon$ always appears 
in this form, so we conclude that $\epsilon=\eta$ is the global
supersymmetric parameter in the AdS space background. 
This spinor 
is known as Killing spinor (see for example \cite{BF}) satisfying the
equation: 
\begin{equation}
 \nabla_{\mu}\eta = -\frac{im}{2}\gamma_{\mu}\eta \quad (D=2, 3),
  \label{killingeq}
\end{equation}
which implies that the supergravity transformation of a gravitino, 
$\delta \psi_{\mu} = 0$.

Here we make some comments on $D=2$ and $3$, $N=1$ supergravity 
multiplet (${e_\mu}^a, \psi_\mu,S$), where ${e_\mu}^a$ is the 
zweibein (dreibein) or \lq\lq graviton\rq\rq,
$\psi_\mu$ is the Rarita-Schwinger field or \lq\lq gravitino\rq\rq, and
$S$ is the auxiliary scalar field, transforms under an infinitesimal
parameter $\epsilon$ as \cite{2DSG}
\begin{eqnarray}
&&\delta{e_\mu}^a={\bar\epsilon}\gamma^a\psi_\mu,\
\delta\psi_\mu=2(\partial_\mu+\frac{1}{2}\omega_\mu\gamma_5+\frac{i}{2}
\gamma_\mu S)\epsilon,\
\delta S=-\frac{1}{2}S{\bar\epsilon}\gamma^\mu\psi_\mu
+\frac{i}{2}e^{-1}\epsilon^{\mu\nu}{\bar\epsilon}\gamma_5\psi_{\mu\nu}
\nonumber\\
&&\hspace{12cm}(D=2),\\
&&\delta{e_\mu}^a={\bar\epsilon}\gamma^a\psi_\mu,\
\delta\psi_\mu=2(\partial_\mu+\frac{1}{2}{\omega_\mu}^c\gamma_c+
\frac{i}{4}\gamma_\mu S)\epsilon,\
\delta S=-\frac{1}{2}S{\bar\epsilon}\gamma^\mu\psi_\mu
+\frac{i}{2}e^{-1}\epsilon^{\mu\nu\lambda}{\bar\epsilon}\gamma_\lambda
\psi_{\mu\nu}
\nonumber\\
&&\hspace{12cm}(D=3),
\end{eqnarray}
where 
$\displaystyle{\psi_{\mu\nu}\equiv \nabla_\mu\psi_\nu-\nabla_\nu\psi_\mu
=(\partial_\mu+\frac{1}{2}\omega_\mu\gamma_5)\psi_\nu-
(\mu\leftrightarrow\nu)}$ for $D=2$, with 
$\displaystyle{\omega_\mu\gamma_5}$ repalced by $\displaystyle{
{\omega_\mu}^c\gamma_c}$ for $D=3$.
To realize the global AdS supersymmetry, we take the supergravity
multiplet to the AdS fixed background: \cite{ST,IO} 
\begin{equation}
{e_\mu}^a={e_\mu}^a({\rm AdS}),\quad \psi_\mu=0,\quad
 S=m\ (D=2),\ 2m \ (D=3)\label{adsbgeq}
\end{equation}

Now let us go back to
the transformation laws for the component fields given as
\begin{eqnarray}
 \delta A &=& \baeta\psi , \nonumber \\
 \delta\psi &=& (F-i\gamma^{\mu}\partial_{\mu}A)\eta ,\nonumber \\
 \delta F &=& -i(\baeta\gamma^{\mu}\nabla_{\mu}\psi)
	\quad (D=2) \\
          &=& -i(\baeta\gamma^{\mu}\nabla_{\mu}\psi)
	 -\frac{m}{2}(\baeta\psi)
	\quad (D=3). \nonumber
\end{eqnarray}
These are in agreement with the fixed background result
for the supergravity transformation \cite{2DSG}.

To construct an invariant action, we also need the invariant
volume element, which is given by
\begin{eqnarray}
 \mathrm{sdet}W_{M}^{\>\>A}
  &=& e\left(1 + \frac{m}{2}(\btheta\theta)\right) \quad (D=2) \nonumber \\
  &=& e\left(1 + m(\btheta\theta)\right) \quad (D=3),
\qquad e = \mathrm{det}e^{a}_{\mu}.
  \label{ospinvariantvolume}
\end{eqnarray}

The invariant action is obtaind as a
generalization of the one in a flat case. For a scalar superfield $\Phi$,
the Kinetic term is obtained as
\begin{equation}
 \mathcal{L}(x)
  = \int\,d^{2}\theta (\mathrm{sdet}W)
  \frac{1}{4}{\bar D^{\alpha}\Phi} D_{\alpha}\Phi. 
\end{equation}
In component fields, this leads to
\begin{equation}
 \mathcal{L}(x)
  = e\left(\frac{1}{2}\nabla^{\mu}A\nabla_{\mu}A
  + \frac{i}{2}(\bpsi\gamma^{\mu}\nabla_{\mu}\psi)
  + \frac{1}{2}F^{2}\right).
   \label{ospkinelag}
\end{equation}

\vspace{0.3cm}
\leftline{\large\bf 4. Nonlinear Transformation Laws and Effective Lagrangians}
\vspace{0.3cm}

In this section we consider the nonlinear transformation laws
and effective lagrangians for NG fields for partial breaking of $N$=2
AdS supersymmetry using the nonlinear
realization method \cite{CWZOg}. 
The relevant coset representive is given by
\begin{equation}
 L(x) = e^{iz(x)\cdot P}e^{i\btheta Q^{1}}
e^{i\bchi(x^{\mu},\theta) Q^{2}}e^{iv(x^{\mu},\theta) T}.
  \label{parl}
\end{equation}
Here we note that the coset space is parametrized by the $N$=1
AdS superspace coordinates $(x^{\mu},\theta)$ as well as by the NG 
superfields $\chi_{\alpha}(x^{\mu},\theta)$ and
$v(x^{\mu},\theta)$, the relevant component of which correspond
to NG fields discussed in the previous section.
Cartan 1-form can be evaluated as
\begin{eqnarray}
 L^{-1}\,dL &=& e^{-ivT}e^{-i\bchi Q^{2}}e^{-i\btheta Q^{1}}e^{-iz(x)\cdot P}
  d\left(e^{iz(x)\cdot P}e^{i\btheta Q^{1}}e^{i\bchi Q^{2}}e^{ivT}\right)
 \nonumber \\
 &\equiv& i\CD x^{a}P_{a} + i\CD\btheta^{\alpha}Q^{1}_{\alpha}
 + i\CD\bchi^{\alpha}Q^{2}_{\alpha} + i\CD vT
 + \frac{i}{2}\mathit{\Delta \tilde{\Omega}}^{ab}M_{ab},
 \label{parcartandef}
\end{eqnarray}
where 
\begin{eqnarray}
 \CD x^{a} &=&
	 Dx^{a}\left(1+\frac{m}{2}(\bchi\chi)\right)+ i(d\bchi\gamma^{a}\chi)
	\quad (D=2) \nonumber \\
	&=& 
	Dx^{a}\left(1+\frac{m}{2}(\bchi\chi)\right)+ i(d\bchi\gamma^{a}\chi)
	 -\frac{1}{4}\mathit{\Delta\Omega}^{bc}\epsilon_{bca}(\bchi\chi)
	\quad (D=3), \nonumber \\
 \CD \btheta^{\alpha}
  &=& i\frac{m}{2}Dx^{a}(\bchi\gamma_{a})^{\alpha}\sin{(mv)}
  + \left(1+\frac{m}{4}(\bchi\chi)\right)D\btheta^{\alpha}\cos{(mv)} 
  \nonumber \\
 & & \quad {} - \frac{1}{4}\mathit{\Delta}\Omega^{ab}\epsilon_{ab}
  (\bchi\gamma_{5})^{\alpha}\sin{(mv)}
  - \left(1-\frac{m}{4}(\bchi\chi)\right)d\bchi^{\alpha}\sin{(mv)}
   \quad (D=2) \nonumber\\
  &=& i\frac{m}{2}Dx^{a}(\bchi\gamma_{a})^{\alpha}\sin{(mv)}
	+ \left(1+\frac{m}{2}(\bchi\chi)\right)D\btheta^{\alpha}\cos{(mv)}
	\nonumber \\
  & & \quad {} + \frac{i}{4}\mathit{\Delta}\Omega^{ab}
  (\bchi\sigma_{ab})^{\alpha}\sin{(mv)}
	- \left(1-\frac{m}{2}(\bchi\chi)\right)d\bchi^{\alpha}\sin{(mv)}
	\quad (D=3), \nonumber \\
  \CD v &=& -(D\btheta\chi) + dv \quad (D=2) \nonumber \\
	&=& -2(D\btheta\chi) + dv \quad (D=3). 
\end{eqnarray}
\relax From these we obtain the superzweibein (dreibein) modified by NG fields
in the similar manner of $N$=1 case,
\begin{equation}
 \CD z^{A} = dz^{M}\CW_{M}^{\>\>A}.
  \label{parzweidef}
\end{equation}
The supercovariant derivatives of NG fields are also modified by 
the existence of NG fields and they are defined as
\begin{equation}
 \CD_{A}\Phi = \frac{\CD\Phi}{\CD z^{A}} = (\CW^{-1})_{A}^{\>\>M}\frac{\CD\Phi}{dz^{M}},
 \quad
 D\Phi \equiv d\Phi + \frac{i}{2}\mathit{\Delta}\omega^{bc}\Sigma_{bc} 
 \Phi.
\end{equation}

Now we determine the constraints which realize the relationship among 
the NG fields discussed in the previous section. 
In general the constraints should be
invariant under the nonlinear transformations. The modified
supercovariant derivatives of NG fields 
satisfy this condition. So we take the constraint as follows,
\begin{equation}
 \cfrac{\CD v}{\CD \btheta^{\alpha}}
  = (\CW^{-1})_{\alpha}^{\>\>\mu}\cfrac{\CD v}{dx^{\mu}}
  + (\CW^{-1})_{\alpha}^{\>\>\beta}\cfrac{\CD v}{d\btheta^{\beta}}
  = 0.
  \label{parconstraint}
\end{equation}
To see what this constraint means, let us rewrite it by
$N$=1 notations. Here we consider the $D$=2 case.
The $D$=3 case can be treated in the same way. 
\relax From the result 
of the Cartan form,
\begin{equation}
 \CD v = -\left(dx^{\mu}W_{\mu}^{\>\>\alpha}
	   + d\btheta^{\beta}W_{\beta}^{\>\>\alpha}\right)\chi_{\alpha}
          + dv, \label{eq:CWdef}
\end{equation}
so the constraint is
\begin{eqnarray}
  \cfrac{\CD v}{\CD \btheta^{\alpha}}
  &=& -\left((\CW^{-1})_{\alpha}^{\>\>\mu}W_{\mu}^{\>\>\gamma}
	+ (\CW^{-1})_{\alpha}^{\>\>\beta}W_{\gamma}^{\>\>\gamma}
	\right)\chi_{\gamma}
   + (\CW^{-1})_{\alpha}^{\>\>\mu}\cfrac{\partial v}{\partial x^{\mu}}
   + (\CW^{-1})_{\alpha}^{\>\>\beta}\cfrac{\partial
  v}{\partial\btheta^{\beta}}
   \nonumber \\
  &=& 0. \label{parconstraint2}
\end{eqnarray}
\relax From (\ref{eq:CWdef}) we obtain
$(\CW^{-1} W)_{A}^{\>\>B} = \delta_{A}^{\>\>B} + \mathcal{O}((v,\chi)^{2})$ and 
the constraint becomes
\begin{eqnarray}
 \chi_{\alpha} &=&  \cfrac{Dv}{D\btheta^{\alpha}} + \mathcal{O}((v,\chi)^{3}) \quad (D=2)
	\nonumber \\
	&=& \frac{1}{2}\cfrac{Dv}{D\btheta^{\alpha}}
	 + \mathcal{O}((v,\chi)^{3}) \quad (D=3).
  \label{chieqv}
\end{eqnarray}
This means that the NG ferimon corresponds to the broken $Q^{2}$, which 
is the superpartner of NG boson corresponding to the broken $T$ and we can
express the NG fields in terms of a scalar superfield $v$.

Now we evaluate nonlinear transformation laws for
$x^{\mu},\theta, v, \chi$ under broken symmetries, $Q^{2}$, $T$.
The procedure is similar to the way we obtain $N$=1 supersymmetric
transformation. The $T$ transformation law turns out to be as follows
\begin{eqnarray}
 \delta x^{\mu} &=& 0, \nonumber \\
 \delta\btheta^{\alpha} &=& m\alpha(1-\frac{m}{4}\btheta\theta)
	\bchi^{\alpha}\quad (D=2)\nonumber \\
	&=& m\alpha(1-\frac{m}{2}\btheta\theta)
	\bchi^{\alpha}\quad (D=3),\nonumber \\
 \delta\bchi^{\alpha} &=& -m\alpha(1+\frac{m}{4}\bchi\chi)
	\btheta^{\alpha}\quad (D=2)\nonumber \\
	&=& -m\alpha(1+\frac{m}{2}\bchi\chi)
	\btheta^{\alpha}\quad (D=3),\label{ttrans}\\
 \delta v &=& \alpha\left( 1 - \frac{m}{2}\btheta\theta\right)
	\left(1- \frac{m}{2}\bchi\chi\right) \quad (D=2) \nonumber \\
	  &=& \alpha\left(1 - m\btheta\theta\right)
	\left(1- m\bchi\chi\right) \quad (D=3). \nonumber
\end{eqnarray}
Similarly, the lowest-order transformation laws under the $Q^{2}$
 transformation are
\begin{eqnarray}
 &\delta x^{\mu} = i(\baeta\gamma^{\mu}\chi) + \cdots,\ \ 
 \delta\btheta^{\alpha} = 0 + \cdots, \ \ 
 \delta\bchi^{\alpha} = \baeta^{\alpha} + \cdots&, \nonumber \\ 
 &\delta v = (\baeta\theta) + \cdots \quad (D=2),\ \ 
	= 2(\baeta\theta) + \cdots \quad (D=3).&
\end{eqnarray}

We now consider the effective Lagrangian for the NG multiplet.
In the spirit of a nonlinear realization method, the invariant
phase volume $\mathrm{sdet}\CW$ is the simplest and
is possible to give the full Lagrangian as used in \cite{Z}. 
But in this case, unfortunately, this does not give rise to the correct
kinetic terms: the ghost appears. This was expected because the
ghost appears also in the case of the partially broken 
super-Poincar{\' e} symmetry \cite{SW}.
So one has to construct
the Lagrangian using the nonlinear transformation laws according 
to \cite{BG1}. That is to say, we write down the general $N$=1
invariant Lagrangian and examine the condition for 
this Lagrangian to be invariant under $Q^{2}$ and $T$
transformations. 
If the Lagrangian is invariant under $T$ and $Q^{1}$,
the Jacobi identity:
\begin{equation}
im\ [Q^{2}, \mathcal{L}] =[T,[Q^{1},\mathcal{L}]]
-[Q^{1},[T,\mathcal{L}]]
\end{equation}
leads to the invariance under $Q^{2}$. Thus we only
have to consider the $T$ transformation.

Here we consider the second order of fields of the Lagrangian
and find the existence of the characteristic ``mass'' term 
for the NG fields.

The general $N$=1 Lagrangian can be represented to the second
order of the fields, both for $D$=2 and $D$=3 cases,
\begin{equation}
 \mathcal{L}(x) = \int\!d^{2}\theta\,
  \mathrm{sdet}W
  \left(\frac{1}{4}\overline{D^{\alpha}}vD_{\alpha}v 
   + \frac{\mu}{2} v^{2}\right), \label{laglag}
\end{equation}
where $\mu$ is a mass parameter to be determined. 
Under the $T$ transformation,
the variation is
\begin{equation}
 \delta\mathcal{L} = \int\!d^{2}\theta\,
	\mathrm{sdet}W
  \left(\frac{1}{2}\overline{D^{\alpha}}vD_{\alpha}\delta v
   + \mu v\delta v\right).
   \label{parlagc}
\end{equation}
Using (\ref{ttrans}) this leads to
\begin{eqnarray}
 \delta\mathcal{L}
  &=& 2\left(-\frac{\alpha m}{2} + \frac{\mu}{2}\alpha \right)F \quad (D=2) \\
  &=& 2\left(-\alpha m + \frac{\mu}{2}\alpha \right)F \quad (D=3). \nonumber
\end{eqnarray}
Therefore if
\begin{equation}
 \mu = m \quad(D=2),\quad \mu=2m \quad(D=3),
	\label{valuec}
\end{equation}
we are lead to the Lagrangain invariant under $T$ (up to
second order of the fields). 
In fact we confirm that this is also invariant under $Q^{2}$ by
using the property of the Killing spinor (\ref{killingeq}). 

In terms of the component fields, the effective Lagrangian (\ref{laglag})
for the NG fields is
\begin{eqnarray}
 e^{-1}\mathcal{L}(x)
 &=& \frac{1}{2}\nabla_{\mu}A\nabla^{\mu}A
  + \frac{i}{2}\bpsi\gamma^{\mu}\nabla_{\mu}\psi + \frac{1}{2}F^{2}
  + mFA - \frac{m}{2}\bpsi\psi + \frac{m^2}{2}A^{2} \quad (D=2)  \\
 &=& \frac{1}{2}\nabla_{\mu}A\nabla^{\mu}A
  + \frac{i}{2}\bpsi\gamma^{\mu}\nabla_{\mu}\psi + \frac{1}{2}F^{2}
  + 2mFA - \frac{3m}{4}\bpsi\psi + 2m^{2}A^{2} \quad (D=3). \nonumber
  \label{componentlag}
\end{eqnarray}
We can eliminate the auxiliary field $F$ using an equation of motion and we get
\begin{eqnarray}
 e^{-1}\mathcal{L}(x)
 &=& \frac{1}{2}\nabla_{\mu}A\nabla^{\mu}A
  + \frac{i}{2}\bpsi\gamma^{\mu}\nabla_{\mu}\psi
  - \frac{m}{2}\bpsi\psi  \quad (D=2) \nonumber  \\
 &=& \frac{1}{2}\nabla_{\mu}A\nabla^{\mu}A
  + \frac{i}{2}\bpsi\gamma^{\mu}\nabla_{\mu}\psi
  - \frac{3m}{4}\bpsi\psi  \quad (D=3).
  \label{componentlagmass}
\end{eqnarray}
Note that in the Lagrangian (\ref{componentlagmass}) the NG fermion $\psi$
possesses a \lq\lq mass\rq\rq\ term both for $D$=2 and 3 with the 
characteristic values depending on the space-time dimension, while the scalar
$A$, which is the NG boson of the broken $T$ does not have the 
\lq\lq mass\rq\rq\  term. 
Here we also note that these mass terms can be derived from the invariant
Lagrangian of the supergravity coupled to matter fields by taking the
the supergravity multiplet to the AdS background (\ref{adsbgeq}).

\vspace{0.3cm}
\leftline{\large\bf 5. Conclusion and Discussion}
\vspace{0.3cm}

In this paper we have investigated the nonlinear realization of 
2 and 3-dimensional $N=2$ AdS supersymmetry which is partially broken
to $N=1$ AdS supersymmetry. We have particularly studied the NG degrees
of freedom, the nonlinear transformation laws and the lowest order 
Lagrangian with a particular attention on the \lq\lq mass\rq\rq\  term. 

Finally, a comment on the 4 dimensional case is in order.
The similar analysis in the present paper can be performed for $D=4$
case. The superfield formalism has already been discussed in \cite{IS}. 
Starting from $N=1$ supergravity, by setting the scalar auxiliary field
$S=3m$, and remaining pseudo-scalar and vector auxiliary fileds to
zero, we arrive at the AdS background: $\delta\psi_\mu=0$ satisfying the
Killing-spinor condition: (\ref{killingeq}).

For the partially broken $N=2$ AdS denoted by $OSp(2,4)$, 
the on-shell degree of freedom corresponding to 
$Q^2$ is two real and the one associated with $T$ is only one because $T^{12}
\equiv T$
is hermitian $SO(2)$ generator. So we need extra one real pseudo-scalar to form
a chiral multiplet or take a vector multiplet and regard the NG boson for $T$
as a auxially field of the vector multiplet. 

\vspace{0.3cm}

{\it Note added}: Recently we have noticed that a nonlinear realization of
$N=2$, $D=3$ Poincar{\'e} supersymmetry down to $N=1$, $D=3$ has been
studied in ref.\cite{EK} to construct the action of $N=1$, $D=4$ 
supermembrane, by contracting the AdS radius.

\newpage
\vspace{0.3cm}
\leftline{\large\bf Acknowledgements}
\vspace{0.3cm}
We thank the members of our High-Energy Theory Group for useful discussions
and encouragements. This work is partially supported by the Grant-in-Aid
for Scientific Research on Priority Area No.707, Japan Ministry of Education.
 

\vspace{1cm}

%

\end{document}